\documentclass{article}
\usepackage{graphicx} 
\usepackage{authblk}
\usepackage{amsmath}
\usepackage{multirow}
\usepackage[sort, comma, numbers]{natbib}
\usepackage{pdflscape}
\usepackage[skip=5pt plus1pt, indent=40pt]{parskip}

\usepackage{geometry}
\usepackage{fancyhdr}
\title{MaskCycleGAN-based Whisper to Normal Speech Conversion}
\author[1]{K. Rohith Gupta}
\author[1]{K. Ramnath}
\author[2]{S. Johanan Joysingh}
\author[1]{\authorcr P. Vijayalakshmi}
\author[3]{T. Nagarajan}
\affil[1]{Sri Sivasubramaniya Nadar College of Engineering, Chennai}
\affil[2]{Vellore Institute of Technology, Chennai}
\affil[3]{Shiv Nadar University Chennai}
\date{}

\begin{document}

\maketitle

\begin{abstract}
Whisper to normal speech conversion is an active area of research. 
Various architectures based on generative adversarial networks have been proposed in the recent past. 
Especially, recent study shows that MaskCycleGAN, which is a mask guided, and cyclic consistency keeping, generative adversarial network, performs really well for voice conversion from spectrogram representations.
In the current work we present a MaskCycleGAN approach for the conversion of whispered speech to normal speech.
We find that tuning the mask parameters, and pre-processing the signal with a voice activity detector provides superior performance when compared to the existing approach.
The wTIMIT dataset is used for evaluation. 
Objective metrics such as PESQ and G-Loss are used to evaluate the converted speech, along with subjective evaluation using mean opinion score. 
The results show that the proposed approach offers considerable benefits. 
\end{abstract}

\vspace{0.25cm}
\begin{center}
\end{center}

\section{Introduction}
\label{sec:intro}
\begin{figure}
    \centering
    \includegraphics[width=0.90\textwidth]{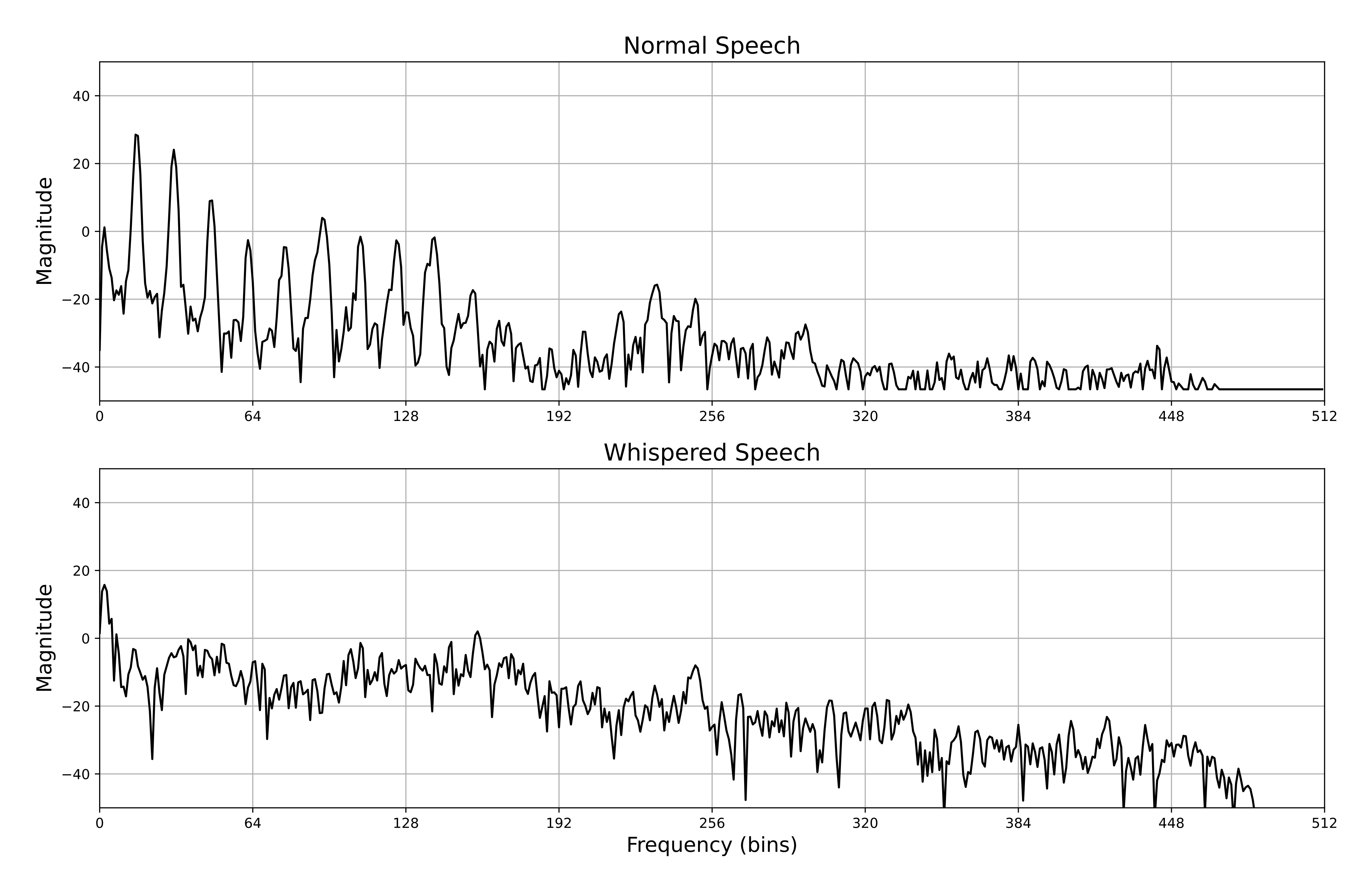}
    \caption{Difference between the spectral envelope of normal and whispered speech}
    \label{fig:spectrum-normalvswhisper}
\end{figure}

Human speech consists of voiced and unvoiced sounds.
Whether a sound is voiced or unvoiced is mainly dependent on the source of excitation, which here is the air flow from the lungs through the glottis. 
When the air flow is modulated by the vibrations of glottis, producing pulses of air flow, it produces voiced sounds. 
This periodicity is reflected as the pitch of the voiced sound.
In contrast when the airflow is not pulsed, it produces unvoiced sounds. 
Whisper is a mode of human speech communication in which only the later excitation exists.
Even the voiced sounds in whisper are produced the same way. 
Figure \ref{fig:spectrum-normalvswhisper} shows the spectrum of normal and whispered speech.
The presence of pitch harmonics in normal speech, and its absence in whispered speech, can be readily observed.

Whisper as a mode of human computer interaction is more common than ever. 
While the need for whisper versus normal speech classification, whisper speech recognition, and whisper speech synthesis, are readily understandable and accepted, there is also a need for whisper to normal speech conversion, and vice-versa. 
This is because while it maybe required for a person to whisper from one end of a communication channel, it may sometimes be more practical for the recipient to receive it as normal speech.
This opens up possibilities such as attending an online meeting or seminar from a quiet environment while someone else in the room is asleep, making it a necessity to be quiet. 
Here although the speaker has the requirement to whisper, it would be beneficial and practical for the other participants of the meeting to receive their voice as normal speech.   
Besides having such luxuries, it should also be noted that thousands of people who have had laryngectomees can only whisper \cite{brook2013laryngectomee}.
Hence, whisper to normal speech conversion serves as a non-invasive solution towards inclusivity. 

\subsection{Whisper to Normal Speech Conversion in Literature}
Whisper to normal speech conversion is carried out mainly in two ways.
The first method fully relies on the source-system paradigm of speech, and can be called rule-based methods.
Here the whisper excitation is replaced with a synthesized voiced excitation where ever required, and the parameters relating to the vocal tract are also sufficiently modified to account for the differences between whisper and normal speech. 
Although this method sounds intuitive, it has its limitations because aspects such as generating a high quality voiced source that preserves naturalness, estimating the voiced and unvoiced regions, incorporating characteristics like the pitch contour, and adapting the vocal tract, are not trivial \cite{toda2012statistical}.

The second method relies on a parallel database of whispered and normal speech, which are used to train a machine learning algorithm.
The main function of the machine learning algorithm here is to map the spectral characteristics of whisper to those extracted from normal speech.
The excitation signals are then generated and its spectrum shaped by an appropriate filter, to produce normal speech.
The conversion can be said to be carried out by whisper feature adaptation, and feature prediction \cite{mcloughlin2017speech}. 

\subsection{Mask Cycle-Generative Adversarial Networks}
The use of a GAN for whisper to normal speech conversion can be found in \cite{pascual2018whispered}.
This method does not involve any parameterization of speech, as it works on raw spectrograms.
The task of converting whisper to normal speech in the case of GANs, is assumed to be an image (spectrogram) translation task. 
The issue with using a basic GAN is that, since image translation can be an under-constrained problem in most cases, it suffers from mode collapse. 
Mode collapse occurs during training where all the input images lead to the same output image, and optimization fails to make progress. 
To counter this problem, the adversarial loss used in basic GANs is compounded with cycle-consistency loss. 
The basic idea behind the cycle consistency loss is to continuously monitor whether the network which translates from a domain X to a domain Y, also has the capability to cycle back to domain X from domain Y, and vice-versa from domain Y to X to Y. 
It was first introduced in \cite{zhu2017unpaired}, where it was used to perform unpaired image to image translation.
Unpaired meaning that it does not require a parallel dataset. 
The shortcoming of using GAN with cycle-consistency, or Cycle-GAN, as noticed in \cite{hu2019mask} for shadow removal, is that, when training with different shadows for the same image, the network does not know which shadow to learn. 
This means that cycle-consistency will not hold.
To handle this, a mask guided approach is proposed, where the masks provide additional information about the shape of the shadow dealt with in the current image - the Mask Cycle-GAN.
The masks are binary, and are stored and updated during training.
In the test phase, the most likely mask is picked.

A series of research works based on Cycle-GAN and MaskCycle-GAN can be seen in \cite{kaneko2019cyclegan}, \cite{kaneko2020cyclegan}, and \cite{kaneko2021maskcyclegan}, for the voice conversion task.
In \cite{kaneko2021maskcyclegan}, the authors propose MaskCycleGAN-VC, which learns the patterns from Mel-spectrograms using a novel self-supervised auxiliary training task named filling in frames (FIF).
It is based on CycleGAN-VC2 \cite{kaneko2019cyclegan}, and is proposed as an alternative to, and improvement over CycleGAN-VC3 \cite{kaneko2020cyclegan}, since CycleGAN-VC3 needs more parameters for training.  
MaskCycleGAN-VC acts as the baseline for the current work.
The masking procedure here is different from that of \cite{hu2019mask}.
Here, during training, masking is used to hide spectral information in order for the network to learn to reproduce it.
For testing, a null-effect all-ones mask is used. 

\section{Proposed Method}
\label{sec:proposed}
In the current work, we propose the Mask Cycle-GAN for the conversion of whisper to normal speech. 
The conversion is considered as an image translation task, wherein the spectrogram of whisper is converted to that of normal speech. 
As seen in Section \ref{sec:intro}, the difference between whisper and normal speech can be mainly attributed to the presence of pitch harmonics in normal speech.
Hence, the task of the proposed system would be: 
(i) to learn to incorporate pitch harmonics in whispered speech, and 
(ii) to learn to incorporate it only in the voiced regions.

Two image processing tasks, that Cycle-GANs are good at, that are of particular interest to the current task as well, are collection style transfer and season transfer \cite{zhu2017unpaired}. 
In the first case, the artistic style (brush strokes, texture, color, etc.) of a particular artist, or collection, is learnt and transferred to another set of test images.
For example, the system would answer the question: ``how would the painter Claude Monet paint the landscape in this image'', through an output image. 
In the second case, an image of a particular location obtained in summer (for example) is provided as the input, and the output is an estimation of how the place would look in the winter.
Thus the overall structure of the image is retained, only changing certain specific attributes of the image, here for example: addition of snow and absence of leaves.
It is obvious how this is quite related to the current task where a task that could be named spectrogram transfer would take place. 
We require the overall structure of the spectrogram to be retained, while making minor modifications in certain parts of the image, to incorporate pitch information.
The intuition behind using a mask-guided approach is that it would address the second task mentioned above. 
That is to learn to incorporate the characteristics of normal speech only in the voiced regions of whispered speech.
Furthermore, since spectral patterns in speech can repeat, masking will offer a way to capture these patterns.

\subsection{Architecture and Training}
The main difference between our proposed model, and MaskCycleGAN-VC \cite{kaneko2021maskcyclegan} was the fine tuning of mask parameters, and the addition of a voice activity detector (VAD) for pre-processing. 
This is apart from the fact that MaskCycleGAN-VC was proposed for voice conversion originally, and not for whisper to normal speech conversion.
The network architecture similar to that used in CycleGAN-VC2 \cite{kaneko2019cyclegan} is adopted in MaskCycleGAN-VC. 
A pretrained MelGAN vocoder \cite{kumar2019melgan} was used to generate the Mel-spectrograms and also synthesize the waveform post-conversion. 
The converter utilized a 2-1-2D CNN \cite{kaneko2019cyclegan}, while the discriminator employed PatchGAN \cite{li2016precomputed}. 
Four losses in total are learnt. 
Three of the losses, that is adversarial, cyclic-consistency, and identity mapping loss are present in \cite{zhu2017unpaired}. 
The fourth loss, namely a second adversarial loss is added in CycleGAN-VC2 to mitigate the statistical averaging induced in the computation of cyclic-consistency loss \cite{kaneko2021maskcyclegan}.


Identical settings as those in CycleGAN-VC3 \cite{kaneko2020cyclegan} were employed for training. 
More information can be obtained from \cite{kaneko2021maskcyclegan}, Section 4.
Similar to the previous CycleGAN-VC \cite{kaneko2019cyclegan, kaneko2020cyclegan} models, no additional data, pre-trained models, or a time alignment procedure were used during training.

\subsection{Tuning Mask Parameters}
Two mask parameters are fine tuned in the current work. 
\begin{enumerate}
    \item Window size, or Number of frames: The number of subsequent or non-subsequent frames selected randomly from the training utterances, in each training iteration. In the baseline, MaskCycleGAN-VC \cite{kaneko2021maskcyclegan}, this is 64 frames. 
    \item Mask size: The percentage of the number of frames that will be masked, expressed in percentage. In the baseline, this is 25\%. 
\end{enumerate}
Tuning the mask parameters like the mask size and the number of frames can help improve the performance of the model while working with multiple speaker data. 
The mask size, and the number of frames were modified as per Table \ref{tab:mask_parameters}.
The window consisted of non-subsequent frames, as per one of the $\mathrm{FIF_{NS}}$ case in \cite{kaneko2021maskcyclegan}. 
The training was carried out for 400 epochs.
This reduced the generator loss, 
which indicated that the generator was able to create a spectrogram that closely corresponds to the speaker’s normal voice.

\begin{table}[]
    \centering
    \caption{Tuning Mask Parameters. The mask size and the number of frames in the baseline, and the proposed systems.}
    \label{tab:mask_parameters}
    \vspace{0.25cm}
    \begin{tabular}{l|c|c}
        \hline
        & \textbf{Baseline} & \textbf{Proposed} \\ \hline
        Number of Frames &  64 & 128 \\
        Mask Size & 25\% & 50\% \\ 
        \hline
    \end{tabular}
\end{table}

\subsection{Voice Activity Detection}
As discussed in Section \ref{sec:proposed}, the advantage of using a mask is that it helps retain the voice characteristics of the speaker, and apply normal speech characteristics at the right places in the spectrogram. 
That is, the mask needs to be applied in the voiced parts of the signal for spectral reconstruction.
During development, it was noticed that the utterances had a silence region at the start and the end. 
During the learning or training process it is possible that the mask might be applied on these silent regions, considering the lack of spectral content to be similar to, for example, the tail end of a plosive.
In order to avoid this problem, voice activity detection (VAD) was used to apply the mask only in the regions containing speech.

The voice activity detector is based on the ratio of energy in speech band and the total energy \cite{pang2017spectrum}, and zero crossing rate. 
When zero-crossing rate is high, the frame is considered to be unvoiced and if it is low, the frame is considered to be voiced frame. 
The threshold is set based on empirical evidence. 
For a 10ms sample of clean speech, the zero-crossing rate is approximately 12 for voiced speech, and 50 for un-voiced speech. 
The frame size and frame rate used here is same as that of the spectrogram.
The ratio between energy of speech band and total energy for window was calculated. 
If ratio is more than the threshold the window is labelled as speech.
Whereas the threshold is set to 0.6 for a normal speech, the threshold for whispered speech was set to 0.2. 
Finally a  median filter with length of 0.5s was applied to smooth the detected speech regions.

\section{Experimental Setup and Results}
The experimental setup, the results, and the observations are detailed in the current section.

\subsection{Dataset}
The wTIMIT is used for analysis and evaluation, in the current work. 
The details of the dataset are provided in Table \ref{tab:dataset}.
The wTIMIT is a large dataset that was produced mainly for evaluating whisper to normal speech conversions, and vice-versa. 
The utterances are predominantly parallel. 
The recordings were obtained in two phases, one in Singapore and one in the United States (US). 
In the current work, we confine our evaluations to the recordings obtained from the US. 

\begin{table}[ht!]
\begin{center}
\caption{The number of utterances and the total duration of the dataset considered in the current work.}
\label{tab:dataset}
\vspace{0.25cm}
\begin{tabular}{lcccc}
\hline \hline
 & \multicolumn{2}{c}{Train} & \multicolumn{2}{c}{Test} \\ \cline{2-5}
 & \multicolumn{1}{l}{Normal} & Whisper & Normal & Whisper \\ \hline 
\multicolumn{5}{l}{\textit{wTIMIT}} \\ \hline 
\multicolumn{1}{c}{Num. Utt.} & 10,934 & 10,245 & 725 & 723 \\
 Duration & \multicolumn{1}{l}{15h 31m} & 16h 25m & 1h 19m & 1h 28m \\ \hline \hline
\end{tabular}
\end{center}
\end{table}

\subsection{Experimental Setup}
\begin{figure}
    \centering
    \includegraphics[width=0.90\textwidth]{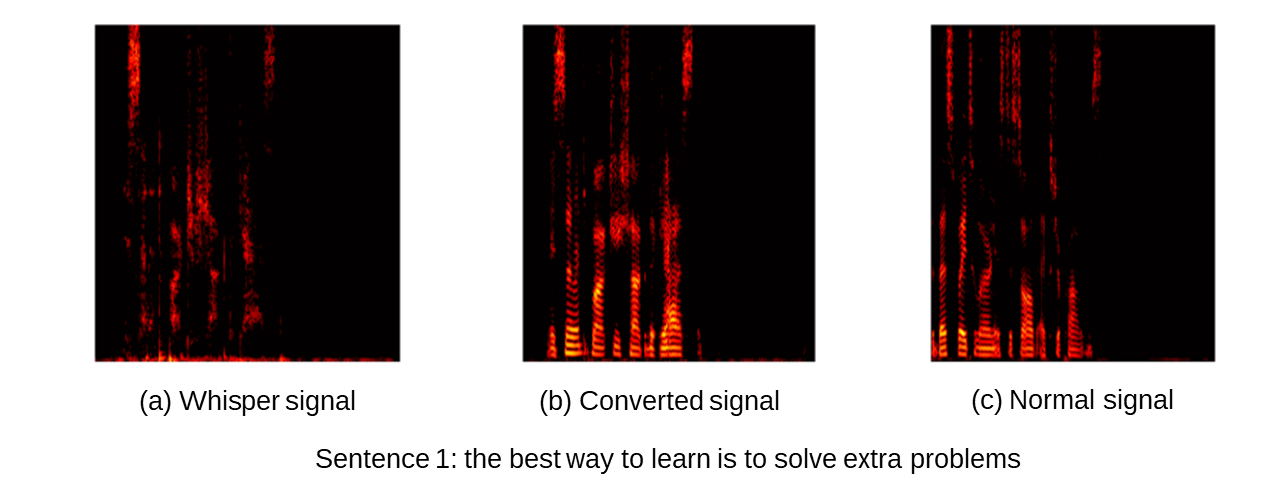}
    \caption{Spectrogram of (a) Whisper (b) Converted (c) Normal speech utterance, after applying VAD}
    \label{fig:after_vad}
\end{figure}

The train and test set are shown in Table \ref{tab:dataset}.
There are 24 speakers which includes both male and female speakers. 
Each speaker had 430 utterances on an average, corresponding to each of normal and whispered speech.
The whisper to normal conversion is speaker dependent, hence the training and testing are carried out one speaker at a time.
To generate the spectrograms initially, the frame size was set to 20ms, and the frame shift was set to 5ms. 
After converting all the audio samples to their corresponding spectrograms, the images are resized to a 224x224 pixels. 
The models were trained for 400 epochs.
From Figure \ref{fig:after_vad} we can visually see the similarity between converted signal and normal speech. 
This similarity is evaluated using objective and subjective metrics, and is discussed in detail in the upcoming sections.

\subsection{Objective Metrics}
Evaluation is carried out using two objective evaluation metrics namely, (a) Perceptual Evaluation of Speech Quality (PESQ) \cite{lee2020voice} (b) Generator Loss (G-Loss) \cite{kaneko2021maskcyclegan}, and one subjective metric namely, Mean Opinion Score (MOS).
PESQ is a standardized and objective method for measuring the quality of speech in telecommunication systems, particularly in the presence of various kinds of noise and distortion. 
PESQ works by comparing a reference speech signal to a degraded speech signal and assigning a quality score based on the degree of similarity between the two. 
The score ranges from -0.5 (worst quality) to 4.5 (best quality). 
PESQ takes into account both subjective and objective factors, such as the audibility of speech, the presence of distortion and noise, and the overall naturalness of speech.

In a Generative Adversarial Network (GAN), the generator loss is the objective function that the generator network tries to minimize during training. Mathematically, the generator loss can be written as:
\begin{equation}
    L_G = -log (D(G(x)))
\end{equation}
Where $G$ is the generator which accepts an input image $x$, and $D$ is the discriminator which determines if the image created by the generator is as close to a distribution of real or reference images as possible. 
Intuitively, the generator loss measures how well the generator is able to generate samples that look like real data samples. 
When the discriminator can differentiate between the fake and real data samples with ease, the generator loss will be higher. 
Conversely, if the created samples are indistinguishable from real data samples, the generator loss will be lower. 

Mean Opinion Score (MOS) \cite{adhilaksono2022study} is used to measure the subjective quality of the converted audio utterances, as perceived by human listeners. 
To obtain a MOS score, a group of human evaluators are presented with the converted utterances, and asked to rate it on a scale of 1 to 5, with 1 being the worst quality and 5 being the best quality, based on the audio clarity, and overall quality of speech. 
The ratings from the evaluators are then averaged to obtain a final MOS score for the particular example. 

\subsection{Results and Discussion}
\label{sec:results}

The results of both the objective metrics for the experiments using mask parameters is listed in Table \ref{tab:mask_objective}. 
Bold numbers indicate the best scores. 
From the metric table, it is seen that modifying the mask parameters improves the performance when compared to the baseline model (italicized).
It can also be seen that best results are obtained when VAD is applied. 

\begin{table}[h]
\centering
\caption{PESQ and G-Loss for objective evaluation. The mask length, number of frames, and pre-processing using VAD, are varied.}
\label{tab:mask_objective}
\vspace{0.25cm}
\begin{tabular}{ccccc}
\hline
\textbf{Mask Len.} & \textbf{Num. Frames} & \textbf{VAD} & \textbf{PESQ}  & \textbf{G-Loss} \\ \hline
25\%          & 64      & No & \textit{1.786} & \textit{10.656} \\
25\%          & 128     & No & 2.147 & 8.678  \\
50\%          & 128     & No & 2.328 & 6.432  \\
50\%          & 128     & Yes & \textbf{3.159} & \textbf{5.542}  \\ \hline
\end{tabular}
\end{table}



To obtain the MOS, 43 evaluators were appointed, and 6 converted audio files were selected at random from the test set and presented to them.
The overall MOS obtained for the final system with mask length set to 50\%, number of frames set to 128, with voice activity detection, is 4.1.


\section{Conclusion}
Mask cycle GAN based approach for originally proposed for voice conversion is applied to whisper to normal speech conversion in the current work. 
When the algorithm is fine tuned, and augmented with voice activity detection it produces better performance in terms of objective and subjective measures. 
The experiments show that mask cycle GAN is a valid approach to whisper to normal speech conversion, that the fine tuning of mask parameters is essential, and that pre-processing using voice activity detection offers considerable improvements due to the nature of the masking process.



\end{document}